\documentclass[11pt,twocolumn,twoside]{article}
\usepackage{fully3d}
\usepackage{amsmath,amssymb,amsfonts}
\usepackage{algorithmic}
\usepackage{graphicx}
\usepackage{textcomp}
\usepackage{mathtools}
\newcommand{\norm}[1]{\lVert#1\rVert}

\begin{document}

\title{Improving CT Image Segmentation Accuracy Using \\ StyleGAN Driven Data Augmentation} 

\author[1]{Soham Bhosale}
\author[1]{Arjun Krishna}
\author[2]{Ge Wang}
\author[1]{Klaus Mueller}

\affil[1]{Department of Computer Science,
          Stony Brook University, Stony Brook, NY, USA}
\affil[2]{Biomedical Imaging Center, School of Engineering,
          Rensselaer Polytechnic Institute, Troy, NY, USA}        

\maketitle
\thispagestyle{fancy}


\begin{customabstract}
Medical Image Segmentation is a useful application for medical image analysis including detecting diseases and abnormalities in imaging modalities such as MRI, CT etc. Deep learning has proven to be promising for this task but usually has a low accuracy because of the lack of appropriate publicly available annotated or segmented medical datasets. In addition, the datasets that are available may have a different texture because of different dosage values or scanner properties than the images that need to be segmented. This paper presents a StyleGAN-driven approach for segmenting publicly available large medical datasets by using readily available extremely small annotated datasets in similar modalities. The approach involves augmenting the small segmented dataset and eliminating texture differences between the two datasets. The dataset is augmented by being passed through six different StyleGANs that are trained on six different style images taken from the large non-annotated dataset we want to segment. Specifically, style transfer is used to augment the training dataset. The annotations of the training dataset are hence combined with the textures of the non-annotated dataset to generate new anatomically sound images. The augmented dataset is then used to train a U-Net segmentation network which displays a significant improvement in the segmentation accuracy in segmenting the large non-annotated dataset.  
\end{customabstract}


\section{Introduction}

CT Image Segmentation is a hallmark of Computer-Aided Diagnosis (CAD). Medical professionals often use it to isolate specific areas of interest in a medical image. This allows them to conduct effective medical analyses for detecting abnormalities such as lung cancer and positioning implants \cite{Skourt2018}. However, manual segmentation can become time-consuming and complicated depending on what part of a body is being analyzed. Deep learning can make this task significantly more efficient, affordable, and accessible. Nevertheless, this approach is currently very ineffective due to the low availability of large labeled and annotated training datasets. The reasons for this shortage include privacy concerns and the high cost of labeling CT datasets by human experts.  

In addition, the added challenge with segmenting a CT dataset is that there is a lot of variability between the publicly available datasets. For example, the images frequently have different dosages which is the difference in amount of x-ray radiation used to synthesize CT images or they could be generated from several different machines each having different parameters for CT reconstruction, which for example could mean the thickness of tissue that each image slice represents may vary. Hence, if we train a segmentation network with a dataset that has images of different dosage and scanner properties than the set of images we intend to segment then the deep network will have low accuracy. 

We present a new strategy for augmenting segmented datasets which eliminates the differences between the segmented dataset we use for training and the dataset we wish to segment, by using texture learning. This strategy will create synthetic images which imitate dosage and scanner properties of the unsegmented dataset and maintain anatomical accuracy within the process.  

\begin{figure}
\centerline{\includegraphics[height=5.5cm]{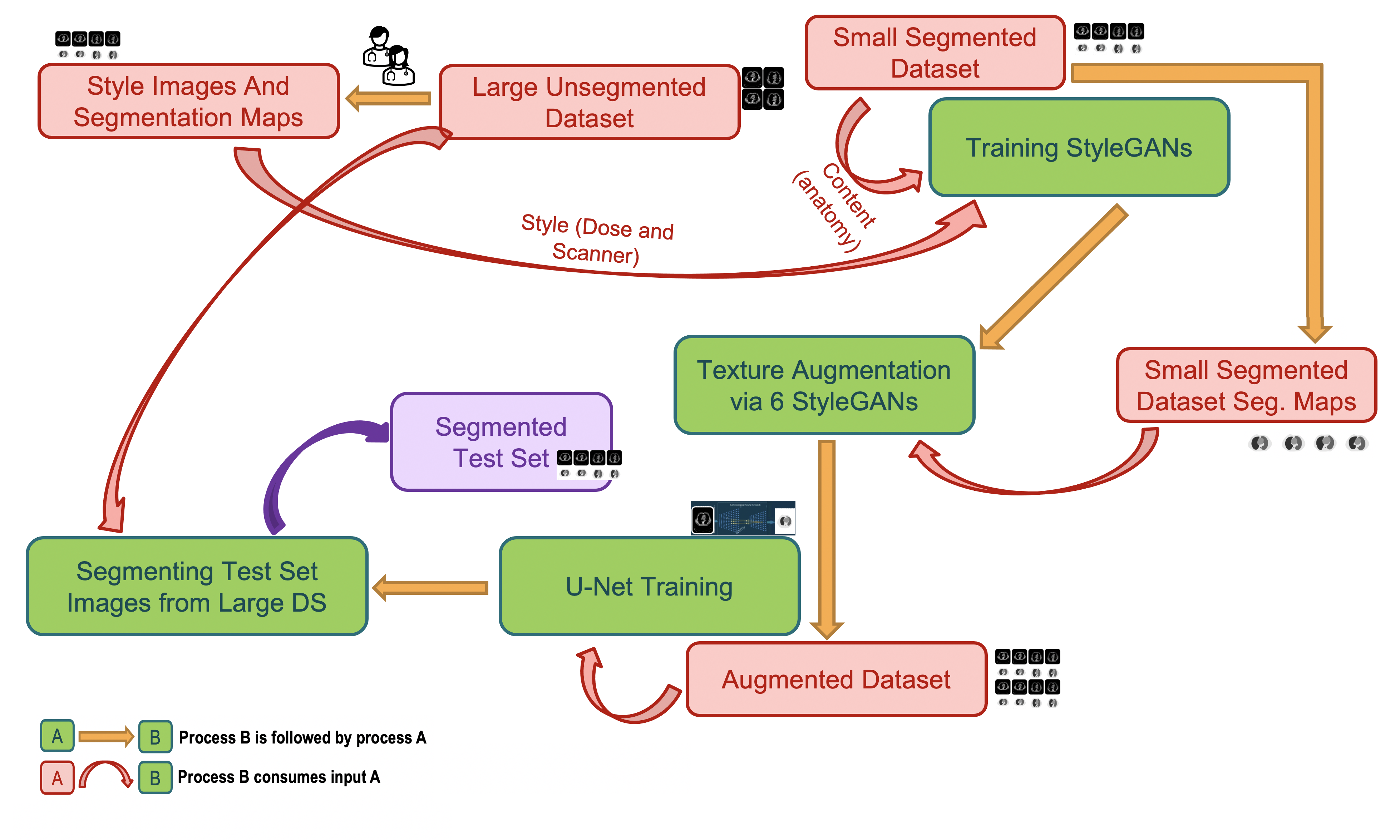}}
\caption{Flow starts at the top right corner with the two datasets -
a small segmented and a large unsegmented dataset. It illustrates the inputs and outputs involved in training the StyleGANs and the U-Net to finally segment the test set taken from the large dataset. }
\label{process diagram}
\vspace{-15pt}
\end{figure}

As shown in Figure 1, we build on our previous work of texture learning \cite{Krishna2019} to expand our small annotated dataset with textures present in the large unlabeled dataset. In the first step, we ask an expert, a radiologist in this case, to manually segment a few CT images from the large non-segmented dataset which were chosen as "style images" for our StyleGAN training. We then use the StyleGAN architecture presented in Krishna et al. \cite{Krishna2019} to create new CT images based on the texture of a style image and the content, or segmentation maps, of another image. 

We train 6 different StyleGANs with 6 separate texture images. Since these texture images came from the non-annotated dataset which contains the images we want to segment, the newly generated segmented images will make good candidates for training the segmentation network. Next, we take all the segmentation maps of the small segmented dataset and use it as "content" inputs to each of 6 trained StyleGAN networks to generate new images. This technique augments the segmented dataset by 7 folds and creates additional images with sample textures from the unsegmented dataset. We then train a U-Net \cite{Ronneberger2015} segmentation network with the augmented segmented dataset. The trained U-Net is then used to segment all the images of the large non-annotated dataset. Segmentation accuracy of the network is computed on a test set that consists of around 100 randomly chosen images from the non-annotated dataset. In the absence of the ground-truth for the test set, annotations are created manually with the assistance of a few radiologists. 

To the best of our knowledge, we are the first to enhance segmentation accuracy by reducing texture differences between training and testing CT datasets using StyleGANs.

\begin{figure}[t!]
\centerline{\includegraphics[height=8.0cm]{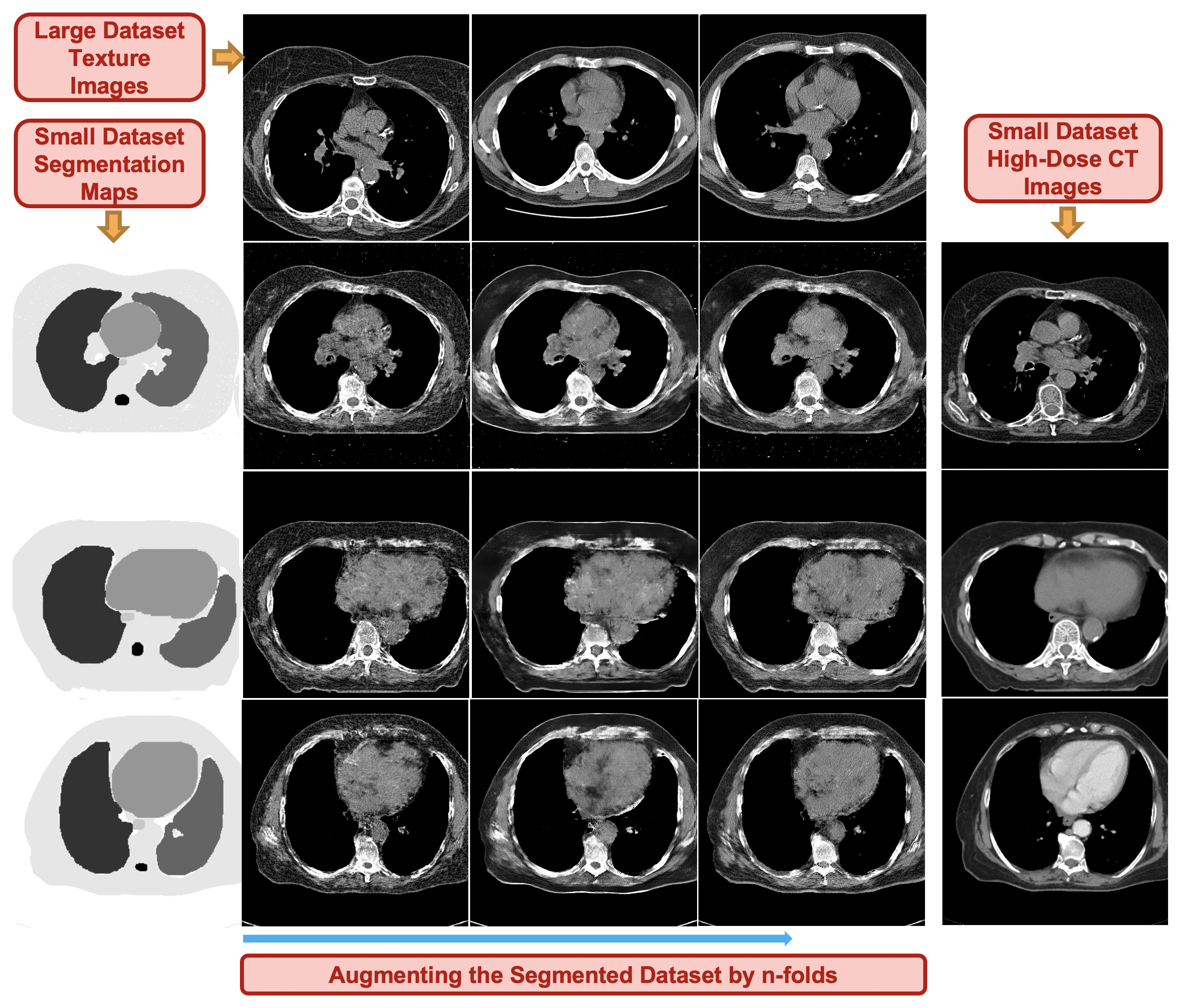}}
\caption{Our Expanded Dataset. For "m" images in annotated dataset and "n" style images, expanded dataset size is mx(n+1). Note, that the heart anatomy isn't perfect in the generated images, but as long as generated images could exhibit enough anatomy and texture, it could still help in improving final segmentation accuracy.}
\vspace{-10pt}
\label{fig}

\end{figure}

\begin{figure*}[t!]
\centerline{\includegraphics[height=3.5cm]{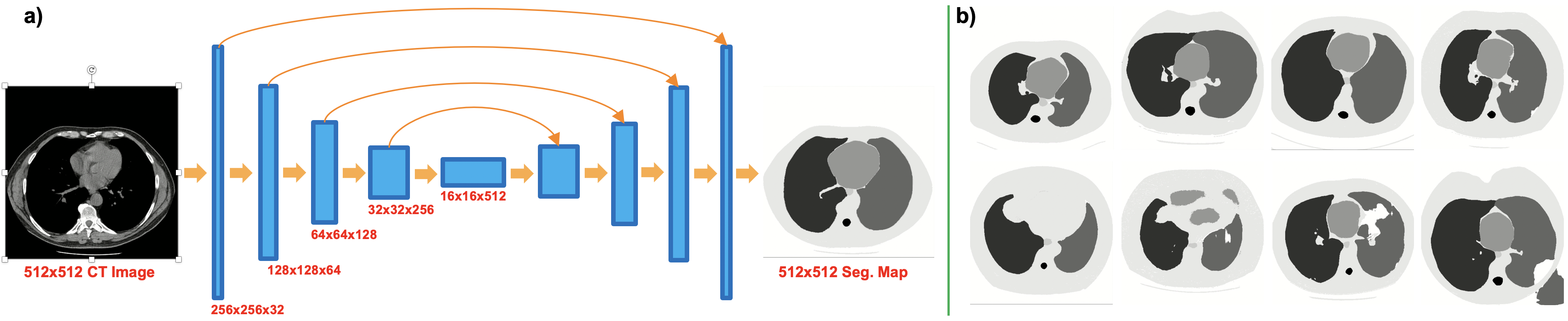}}
\caption{Left: A standard U-Net network which segments a low-dose soft-tissue CT image into lungs, heart, torso, spinal-cord and esophagus. Right: Top row shows some well formed segmentation maps of various ct scans, while bottom row highlights flawed results from the U-Net when trained on a smaller different dataset with high-dose CT images. }
\label{fig}

\end{figure*}

\section{Materials and Methods}

Fig. 1 highlights each step of the process. As discussed, there are three major steps: training the StyleGAN networks, augmenting the annotated dataset, and then training the segmentation U-Net with this augmented dataset. We will briefly describe each step below

\subsection{Datasets}

The top right corner of Fig. 1 highlights the two different datasets (in red colored boxes) that were involved in this work. The smaller dataset consists of 512x512 high dose chest CT images along with their segmentation maps of thirty patients depicting their lungs, heart, spinal cord, esophagus, surrounding tissue and bones within their torso. The larger dataset consists of non-annotated low dose chest CT images of similar resolution of around 14k patients taken with several different scanners at several different locations. As stated before, for using the datasets together for improving segmentation, we pre-process and augment the smaller annotated dataset multiple folds \cite{Krishna2019} using CT image textures present in the larger dataset. 

\subsection{StyleGAN}

We use our encoder-decoder based styleGAN architecture and segment-wise style loss \cite{Krishna2019} for learning and generating segment-wise textures of heart, torso and surrounding tissues present in the larger dataset. For this, we first choose five or six different "style images" from the large dataset. Since the dataset contains low-dose CT scans collected over a period of three years from several different scanners, it contains a few distinct textures over all the CT-scans corresponding to distinct noise characteristics of these scanners. The "style-images" are manually selected corresponding to these distinct textures. For our large dataset six style-images seem enough to cover all the different texture features present in the dataset. Images in the top row of Figure 2 show three such style images. Note that as per our styleGAN architecture, the segmentation maps of both "style images", from the large dataset, and "content images", from the small dataset, are needed for generating CT images having textures of the large dataset's images while retaining anatomy of the small dataset's "content images". This is maintained by our segment-wise style loss, L$_{s}$ and content loss, L$_{s}$ which is as follows:

\vspace{-5pt}

\begin{align}
L_{s} = \sum_{sg{\in}SG}\sum_{l{\in}SL}\frac{1}{4N^{2}_{l}M^{2}_{l}} \norm{ G(R^{l}_{sg}) - G(S^{l}_{sg}) }_{F}^{2}   
\end{align}

\begin{align}
L_{c} = \sum_{sg{\in}SG}\sum_{l{\in}CL}\frac{1}{2N_{l}M_{l}} \norm{ R^{l}_{sg} - O^{l}_{sg} }_{2}^{2}      
\end{align}
where O$^{l}$, S$^{l}$, and R$^{l}$ denote the feature maps extracted from the pre-trained VGG network at layer l, for the original image x$_{o}$, the style image x$_{s}$ and the stylized image output x$_{r}$ respectively; $G(R^{l})$ and $G(S^{l})$  denote the encoded Graham Matrices \cite{gatys2015texture} of those feature maps; SG is the set of all six segments including heart, torso, lungs, spinal cord and esophagus in the CT images and SL/CL is the set of all style/content layers in the VGG network. Since, the large training dataset's CT scans are not segmented, we manually build the segmentation maps of the chosen six style images with the assistance of a radiologist, as mentioned. 

The StyleGAN architecture proposed by Krishna et al. \cite{Krishna2019} takes the segmentation map of one image x$_{o}$ and the segments' texture of another image x$_{s}$ as input, and outputs a new image x$_{r}$ with the contents (organs) of the segmentation map x$_{o}$ and style of image x$_{s}$. Similar to the original work, all the training images are resized to 512x512 pixels. Each StyleGAN is trained for 100 epochs. For the loss function, the input images, generated images, and style images along with their segmentation maps are fed into a VGG-19 network. The feature maps generated by VGG-19 are used to calculate style loss and content loss. Both content loss and style loss are calculated separately for each segmented region and then combined.   

\subsection{Augmenting Segmentation Training Dataset}

Next, all the segmentation maps of the smaller CT segmentation training dataset are passed through each trained StyleGAN networks as inputs generating new CT images for each segmentation map. Note that we train six style generators corresponding to each style image hence augmenting the dataset seven fold. The original 700 images segmented lung CT dataset is now expanded to around 4900 images. In our case, the generated images have textures simulating low dosage CT images similar to the images of the large dataset we intend to segment. Also, all of the generated images have associated annotations or segmentation maps which are the same as the inputs or the annotations of the original dataset 

Figure 2 shows a few results of the expanded annotated dataset. In the figure we are showing results for three such generators corresponding to three style images shown in the top row and three segmentation maps corresponding to three "content"/training dataset images shown in the right-most column. Since these generators do not serve as our final segmentation networks, the generated images do not have to be anatomically prefect as long as they can learn and exhibit enough anatomy with the large dataset's texture in generated images that could help in training a U-Net effectively in our last step. We chose the soft tissue window of CT scans for texture learning and augmenting the smaller dataset. Soft tissue images exhibit all the anatomical details necessary for learning annotations of the smaller dataset which in our case are segmentation maps having segments depicting heart, lungs, torso esophagus and spinal cord. These segments are clearly distinguishable in the soft tissue window based on the pixel values of the images. Having the ability to learn annotations or segmentation from the augmented dataset will enable the trained U-Net to recognize the CT images' segments from the larger dataset having similar textures.

\subsection{Training the U-Net}

We train a U-Net \cite{Ronneberger2015} for creating a segmentation map as output when given a chest CT image as input. We chose the U-Net architecture due to its excellent performance for segmenting biomedical images, especially when coupled with data augmentation techniques like elastic deformation. The architecture of the network is based on the original work done by the authors where each blue block in Figure 3a consists of two convolutional layers with a batch normalization layer in between those layers and a max-pooling layer as the last layer. In the up-sampling part, max-pooling layers are replaced by the transposed convolutions layer as the first layer followed by the two convolutional layers. As shown in the figure the results of each block in the down-sampling part are concatenated to their corresponding block in the up-sampling half. Kernels of size 3x3 are used for convolutional layers with a stride of 2. We use the augmented textured annotated dataset created in the previous step for training our U-Net. Having similar texture across the two different datasets helps training a segmentation network on the segmented dataset to segment the images of the large non-annotated dataset. We use the trained U-Net to segment randomly chosen 100 CT images of the larger dataset which forms our test set. 

We analyze the improvement of the network trained on the augmented CT image dataset as compared to the network trained on the small high-dose image dataset. In the absence of the ground-truth segmentation maps for the test set, we manually corrected the erroneous predicted segmentation outputs of the network models with the help of a few radiologists. The corrected segmentation outputs of the test set images are then treated as the ground truths for the images. The first row of Figure 3b highlights some of these well formed ground truth segmentation maps, while the second row shows the erroneous predicted segmentation maps from the network trained on the smaller high-dose image dataset. We discuss the improvements and compare the two networks in the next section. Both are trained for 30 epochs with a batch size of 16.     

\begin{figure*}[t!]
\centering
\includegraphics[height=12cm, width=17cm]{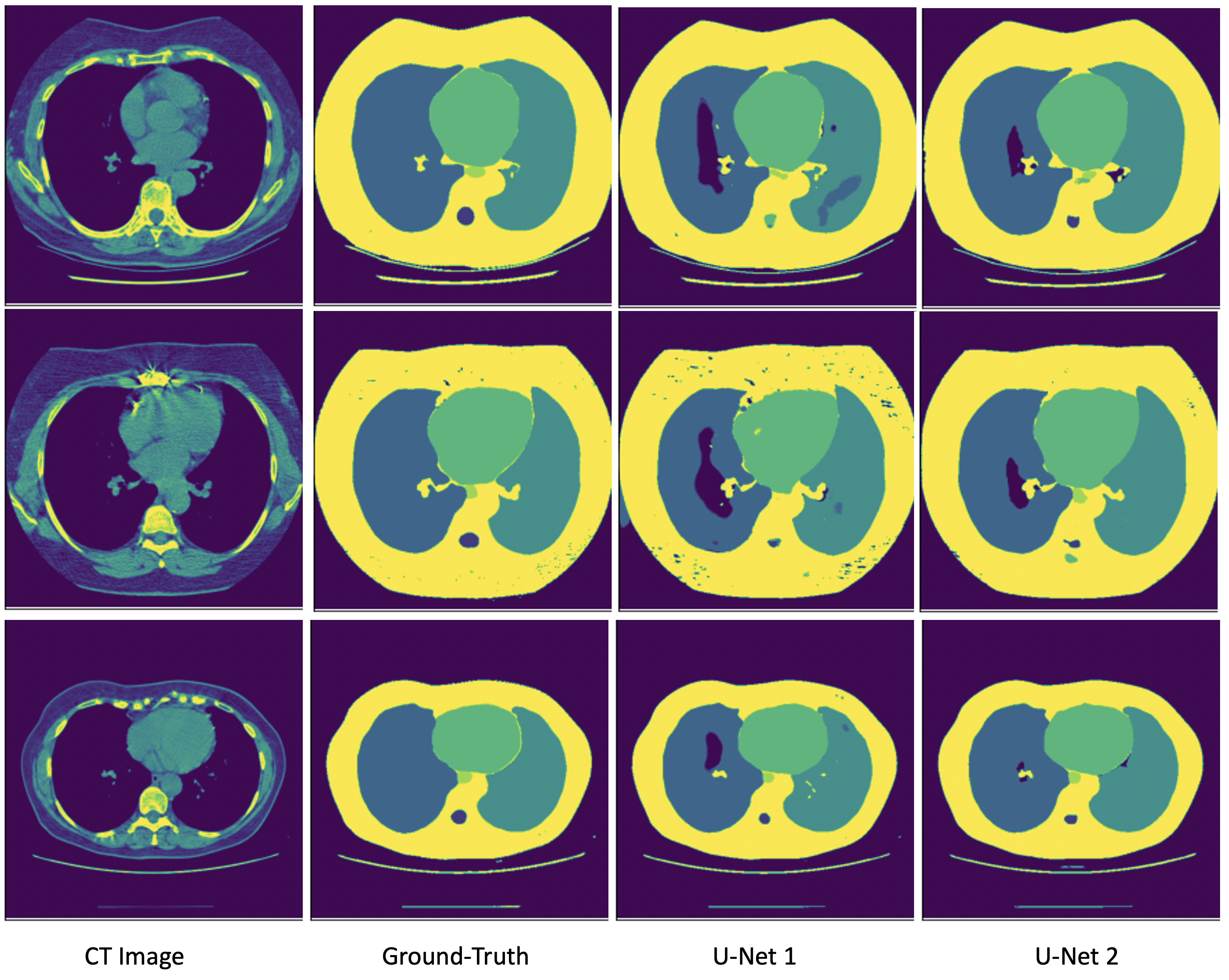}
\caption{The above figures display a side-by-side comparison of the test results of the segmentation network (U-Net 1) trained on the small high-dose CT dataset versus the network (U-Net 2) trained on the augmented dataset. The segmentation maps generated by the augmented dataset show greater visual accuracy than those by the network trained on the small dataset.}
\label{fig}
\vspace{-10pt}
\end{figure*}

\section{Results, Further Work and Conclusion}
Figure 4 highlights some of our results. We observe that the U-Net trained with our style learning based augmented data (U-Net 2) has a greater segmentation accuracy than the network trained with the original annotated dataset (U-Net 1).  For instance, the augmented dataset-trained network classified much more of lung and torso matter correctly than the network trained with the original dataset. As for metrics, U-Net 2 has a testing accuracy of 95.7\% whereas the U-Net 1 has a testing accuracy of 93.1\%. 

We also observe that we still have segmentation flaws around smaller segments like esophagus and spinal cord. Since the small dataset represents limited anatomy of a few patients, there are errors in the segmentation outputs of the patients in the larger dataset with more varied anatomy. This becomes more evident around smaller segments since the segmentation maps of the small dataset are on an average significantly smaller, i.e., have less spatial detail,  than those of the large dataset which is due to the differences in positioning of the scanner lens between the two datasets.  

One way to resolve this is to explore the PCA space of segments within segmentation maps as we did in Krishna el al. \cite{Krishna2019} to generate CT images with varied anatomy through our styleGANs for augmenting the small annotated dataset. Also, we propose to further improve our segmentation network accuracy by training U-Nets and StyleGANs in a cyclic approach. We will use StyleGANs to augment a dataset, pass it through the segmentation network, and then use the segmentation maps generated by this network to provide additional data for the StyleGAN training and then repeat the process. Another option is to use a combination of all CT window images including lung and bone windows for segmenting the images. We propose to use a combination of above mentioned strategies to further improve our segmentation accuracy.   

According to these results, augmenting medical datasets using StyleGANs has proven to be a promising method to resolve texture differences between different medical datasets paving a way to analyze one dataset based on the annotations of another. In the segmentation task of lung CT, the StyleGANs assisted expanded dataset helped a U-Net to improve its accuracy by 2.6\% on a completely different dataset. Hopefully, this new technique can mitigate the annotated medical data shortage issue that is required for training data intensive deep learning networks for various applications.

\printbibliography

\end{document}